\documentclass{emulateapj}

\def\hi{\textsc{Hi} }

\begin{document}

\title{Toward Empirical Constraints on the Global Redshifted 21 cm Brightness Temperature during the Epoch of Reionization}

\author{Judd D. Bowman\altaffilmark{1}, Alan E. E. Rogers\altaffilmark{2}, and Jacqueline N. Hewitt\altaffilmark{1}}

\altaffiltext{1}{Massachusetts Institute of Technology, Kavli
Institute for Astrophysics and Space Research, Cambridge,
Massachusetts, USA}

\altaffiltext{2}{Massachusetts Institute of Technology, Haystack
Observatory, Westford, Massachusetts, USA}

\email{jdbowman@mit.edu}

\begin{abstract}

Preliminary results are presented from a simple, single-antenna
experiment designed to measure the all-sky radio spectrum between 100
and 200~MHz. The system used an internal comparison-switching scheme
to reduce non-smooth instrumental contaminants in the measured
spectrum to 75~mK. From the observations, we place an initial upper
limit of $450$~mK on the relative brightness temperature of the
redshifted 21~cm contribution to the spectrum due to neutral hydrogen
in the intergalactic medium (IGM) during the epoch of reionization,
assuming a rapid transition to a fully ionized IGM at a redshift of
8. With refinement, this technique should be able to distinguish
between slow and fast reionization scenarios.  To constrain the
duration of reionization to $\Delta z>2$, the systematic residuals in
the measured spectrum must be reduced to 3~mK.

\end{abstract}


\section{Introduction}

The transition period at the end of the cosmic ``Dark Ages'' is known
as the epoch of reionization (EOR). During this epoch, radiation from
the very first luminous sources---early stars, galaxies, and
quasars---succeeded in ionizing the neutral hydrogen gas that had
filled the intergalactic medium (IGM) since the recombination event
following the Big Bang. Reionization marks a significant shift in the
evolution of the Universe. For the first time,
gravitationally-collapsed objects exerted substantial feedback on
their environments through electromagnetic radiation, initiating
processes that have dominated the evolution of the visible baryonic
Universe ever since. The epoch of reionization, therefore, can be
considered a dividing line when the relatively simple evolution of
the early Universe gave way to more complicated and more
interconnected processes. Although the Dark Ages are known to end
when the first luminous sources ionized the neutral hydrogen in the
IGM, precisely when this transition occurred remains uncertain.

The best existing constraints on the timing of the reionization epoch
come from two sources: the cosmic microwave background (CMB)
anisotropy and absorption features in the spectra of high-redshift
quasars. The amplitude of the observed temperature anisotropy in the
CMB is affected by Thomson scattering due to electrons along the line
of sight between the surface of last scattering and the detector, and
thus, it is sensitive to the ionization history of the IGM through
the electron column density. In addition, if there is sufficient
optical depth to CMB photons due to free electrons in the IGM after
reionization, some of the angular anisotropy in the unpolarized
intensity can be converted to polarized anisotropy. This produces a
peak in the polarization power spectrum at the angular scale size
equivalent to the horizon at reionization with an amplitude
proportional to the optical depth \citep{1997ApJ...488....1Z}.
Measurements by the WMAP satellite of these effects indicate that the
redshift of reionization is $z_r\approx11\pm4$
\citep{2007ApJS..170..377S}, assuming an instantaneous transition.

Lyman-$\alpha$ absorption by neutral hydrogen is visible in the
spectra of many high-redshift quasars and, thus, offers the second
currently feasible probe of the ionization history of the IGM.
Continuum emission from quasars is redshifted as it travels through
the expanding Universe to the observer.  Neutral hydrogen along the
line of sight creates absorption features in the continuum at
wavelengths corresponding to the local rest-frame wavelength of the
Lyman-$\alpha$ line.  Whereas CMB measurements place an integrated
constraint on reionization, quasar absorption line studies are
capable of probing the ionization history in detail along the
sight-lines. There is a significant limitation to this approach,
however. The Lyman-$\alpha$ absorption saturates at very low
fractions of neutral hydrogen (of order $x_{HI} \approx 10^{-4}$).
Nevertheless, results from these studies have been quite successful
and show that, while the IGM is highly ionized below $z\lesssim6$
(with typical $x_{HI}\lesssim10^{-5}$), a significant amount of
neutral hydrogen is present above, although precisely how much
remains unclear \citep{2001ApJ...560L...5D, 2001AJ....122.2850B,
2002AJ....123.1247F, 2003AJ....125.1649F, 2004Natur.427..815W,
2006AJ....132..117F}.

The existing CMB and quasar absorption measurements are somewhat
contradictory.  Prior to these studies, the reionization epoch was
assumed generally to be quite brief, with the transition from an IGM
filled with fully neutral hydrogen to an IGM filled with highly
ionized hydrogen occurring very rapidly.  These results, however,
open the possibility that the ionization history of the IGM may be
more complicated than previously believed \citep{2003ApJ...595....1H,
2003ApJ...591...12C, 2003MNRAS.344..607S, 2004ApJ...604..484M}.

Direct observations of the 21~cm (1420~MHz) hyperfine transition line
of neutral hydrogen in the IGM during the reionization epoch would
resolve the existing uncertainties and reveal the evolving properties
of the IGM.  The redshifted 21~cm signal should appear as a faint,
diffuse background in radio frequencies below $\nu<200$~MHz for
redshifts above $z>6$ (according to $\nu=1420/[1+z]$~MHz). For
diffuse gas in the high-redshift ($z\approx10$) IGM, the expected
unpolarized differential brightness temperature of the redshifted
21~cm line relative to the pervasive CMB is readily calculable from
basic principles and is given by \citep[their
\S~2]{2004ApJ...608..622Z}
%
\begin{equation}
\begin{array}{rl}
\label{eqn_intro_temp} \delta T_{21}(\vec{\theta}, z) \approx~&
23~(1+\delta)~x_{HI} \left ( 1 - \frac{T_\gamma}{T_S} \right ) \\
& \times \left ( \frac{\Omega_b~h^2}{0.02} \right ) \left [ \left (
\frac{0.15}{\Omega_m~h^2} \right ) \left ( \frac{1+z}{10} \right )
\right ]^{1/2} \mbox{mK},
\end{array}
\end{equation}
%
where $\delta(\vec{\theta},z)$ is the local matter over-density,
$x_{HI}(\vec{\theta},z)$ is the neutral fraction of hydrogen in the
IGM, $T_\gamma(z) = 2.73~(1+z)$~K is the temperature of CMB at the
redshift of interest, $T_S(\vec{\theta},z)$ is the spin temperature
that describes the relative population of the ground and excited
states of the hyperfine transition, and $\Omega_b$ is the baryon
density relative to the critical density, $\Omega_m$ is the total
matter density, and $h$ specifies the Hubble constant according to
$H_0=100~ h$~km~s$^{-1}$~Mpc$^{-1}$. From
Equation~\ref{eqn_intro_temp}, we see that perturbations in the local
density, spin temperature, and neutral fraction of hydrogen in the
IGM would all be revealed as fluctuations in the brightness
temperature of the observed redshifted 21~cm line.

The differential brightness temperature of the redshifted 21~cm line
is very sensitive to the \hi spin temperature. When the spin
temperature is greater than the CMB temperature, the line is visible
in emission. For $T_S \gg T_\gamma$, the magnitude of the emission
saturates to a maximum (redshift-dependent) brightness temperature
that is about 25 to 35~mK for a mean-density, fully neutral IGM
between redshifts 6 and 15, assuming a $\Lambda$CDM cosmology with
$\Omega_m=0.3$, $\Omega_\Lambda=0.7$, $\Omega_b=0.04$, and $h=0.7$.
At the other extreme, when the spin temperature is very small and
$T_S \ll T_\gamma$, the line is visible in absorption against the CMB
with a potentially very large (and negative) relative brightness
temperature.

A number of factors are involved in predicting the typical
differential brightness temperature of the redshifted 21~cm line as a
function of redshift.  In particular, the spin temperature must be
treated in detail, including collisional coupling between the spin
and kinetic temperatures of the gas, absorption of CMB photons, and
heating by ultra-violet radiation from the first luminous sources. We
direct the reader to \citet{2006PhR...433..181F} for a good
introduction to the topic. The results of several efforts to predict
the evolution of the differential brightness temperature of the
redshifted 21~cm line have yielded predictions that are generally
consistent in overall behavior, but vary highly in specific details
\citep{1997ApJ...475..429M, 1999A&A...345..380S, 2004ApJ...608..611G,
2006MNRAS.371..867F}. These models tend to agree that, for a finite
period at sufficiently high redshifts ($z\gtrsim20$), the \hi
hyperfine line should be seen in absorption against the CMB, with
relative brightness temperatures of up to $|\delta
T_b|\lesssim100$~mK.  This is because the IGM initially cools more
rapidly than the CMB following recombination
\citep{1994ApJ...427...25S, 1997ApJ...475..429M}. During this period,
fluctuations in the differential brightness temperature of the
redshifted 21~cm background should track the underlying baryonic
matter density perturbations \citep{1972A&A....20..189S,
1979MNRAS.188..791H, 1990MNRAS.247..510S, 2002ApJ...572L.123I,
2003MNRAS.341...81I, 2004PhRvL..92u1301L, 2005ApJ...626....1B}.
Eventually, however, the models indicate that the radiation from the
first generations of luminous sources will elevate the spin
temperature of neutral hydrogen in the IGM above the CMB temperature
and the redshifted 21~cm line should be detected in emission with
relative brightness temperatures up to the expected maximum values
(of order $25$~mK). Finally, during the reionization epoch, the
neutral hydrogen becomes ionized, leaving little or no gas to produce
the \hi emission, and the apparent differential brightness
temperature of the redshifted 21~cm line falls to zero as
reionization progresses. As the gas is ionized, a unique pattern
should be imprinted in the redshifted 21~cm signal that reflects the
processes responsible for the ionizing photons and that evolves with
redshift as reionization progresses \citep{1997ApJ...475..429M,
2000ApJ...528..597T, 2003ApJ...596....1C, 2004ApJ...608..622Z,
2004ApJ...613...16F}. The details of the specific timing, duration,
and magnitude of these features remains highly variable between
theoretical models due largely to uncertainties about the properties
of the first luminous sources.

Measuring the brightness temperature of the redshifted 21~cm
background could yield information about both the global and the
local properties of the IGM. Determining the average brightness
temperature over a large solid angle as a function of redshift would
eliminate any dependence on local density and temperature
perturbations and constrain the evolution of the product
$\overline{x_{HI}(1-T_\gamma/T_S)}$, where we use the bar to denote a
spatial average. During the reionization epoch, it is, in general,
believed to be a good approximation to assume that $T_S\gg T_\gamma$
and, therefore, that the brightness temperature is proportional
directly to $\bar{x}_{HI}$. Global constraints on the brightness
temperature of the redshifted 21~cm line during the EOR, therefore,
would directly constrain the neutral fraction of hydrogen in the IGM.
Such constraints would provide a basic foundation for understanding
the astrophysics of reionization by setting bounds on the duration of
the epoch, as well as identifying unique features in the ionization
history (for example if reionization occurred in two phases or all at
once).  They would also yield improvements in estimates of the
optical depth to CMB photons and, thus, would help to break existing
degeneracies in CMB measurements between the optical depth and
properties of the primordial matter density power spectrum
\citep{2006PhRvD..74l3507T}. 
\begin{figure}
\centering
\includegraphics[width=20pc]{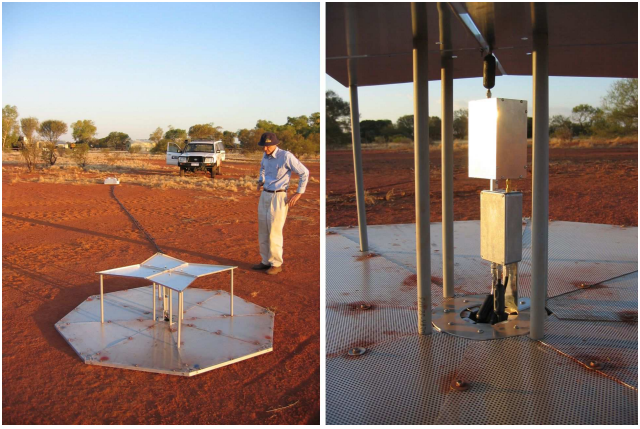}
\caption[Photograph of EDGES deployed at Mileura Station]{
\label{f_edges_photos} EDGES deployed at Mileura Station in Western
Australia. The left panel shows the full antenna and ground screen in
the foreground and the analog-to-digital conversion and data
acquisition module in the background.  The right panel is a close-up
view of the amplifier and switching module connected directly to the
antenna (through the balun).}
\end{figure}
For these reasons, several efforts are underway to make precise
measurements of the radio spectrum below $\nu<200$~MHz ($z>6$).  In
this paper, we report on the initial results of the Experiment to
Detect the Global EOR Signature (EDGES). In \S~\ref{s_edges_method},
we describe the specific approach used for EDGES to address the issue
of separating the redshifted 21 signal from the foreground emission.
We then give an overview of the EDGES system in
\S~\ref{s_edges_system}, followed by the results of the first
observing campaign with the system in \S~\ref{s_edges_results}, along
with a discussion of the implications for future single-antenna
measurements.

\section{Method}
\label{s_edges_method}

In principle, the global brightness temperature measurement is much
less complicated to perform than the detection of local perturbations
in the redshifted 21~cm background (which will be attempted in the
near future by the Mileura Widefield Array [MWA], Low Frequency Array
[LOFAR], Giant Metrewave Radio Telescope [GMRT], and Twenty-one
Centimeter Array [21CMA]). Since the desired signal for the global
measurement is the mean brightness temperature due to redshifted 21
cm emission (or absorption) over the entire sky, there is no need for
high angular resolution or imaging. A single antenna tuned to
the appropriate frequencies could reach the required sensitivity
($\sim25$ mK) within only one hour of integration time, assuming a
reasonable spectral resolution of $\sim1$~MHz (equivalent to $\Delta
z\approx0.1$ at $z\approx8$). There is a fundamental complication
with such an experiment, however, arising from the global nature of
the signal. Since the expected redshifted 21 cm emission fills the
entire sky, there is no ability to perform comparison switching
between the target field and a blank field. The problem this causes
is two-fold. First, it is difficult to separate the contribution to
the measured spectrum due to the redshifted \hi signal from that of
any other all-sky emission, including Galactic synchrotron and
free-free radiation, the integrated contribution of extragalactic
continuum sources, or the CMB. Second, for similar reasons, it is
difficult to avoid confusing any systematic effects in the measured
spectrum due to the instrument or environment with received signal
from the sky. The severity of these problems is exacerbated in
single-antenna measurements by the intensity of the Galactic
synchrotron emission. Unlike interferometric observations, a single
antenna is sensitive to the large-scale emission from the Galaxy,
providing a 200 to 10,000~K foreground in the measured spectrum.

Determining the $\sim25$ mK redshifted 21 cm contribution to the
radio spectrum requires separating the signal from the foreground
spectrum at better than 1 part in 10,000.  This can be accomplished
by taking advantage of the differences between the spectra of the
Galactic and extragalactic foregrounds and the anticipated redshifted
21~cm contribution. As discussed by \citet{1999A&A...345..380S},
Galactic synchrotron emission is the dominate component of the
astrophysical foregrounds below $\nu<200$~MHz, accounting for all but
approximately 30 to 70~K of the foregrounds at 178~MHz
\citep{1967MNRAS.136..219B}. Its spectrum is very nearly a power-law
given, in temperature units, by $T_{gal}(\nu) \sim \nu^{-\beta}$,
where $\beta\approx2.5$ is the spectral index. The spectral index is
generally constant over the frequencies of interest ($50 \lesssim \nu
\lesssim 200$~MHz), although it is known to flatten with decreasing
frequency due to self-absorption. The intensity of the synchrotron
emission and the exact value of the spectral index depend on Galactic
coordinate. The amplitude varies over an order of magnitude, between
about $200 < T_{gal} < 10,000$ K at 150~MHz (peaking toward the
Galactic center and along the Galactic plane), while the spectral
index has small variations of order $\sigma_\beta\approx0.1$
dependent largely on Galactic latitude, with the steepest regions
occurring at high Galactic latitudes. Free-free emission in the
Galaxy and discrete Galactic and extragalactic continuum sources also
have spectra that can be reasonably described by power-laws. The
integrated flux from extragalactic continuum sources is generally
isotropic on large scales and accounts for the majority of the
remaining power in the low-frequency radio spectrum, with free-free
emission making up only about 1\% of the total power. The combined
spectrum due to the astrophysical foregrounds is smooth and remains
similar to a power-law profile.

On the other hand, as the apparent differential brightness
temperature of the redshifted 21~cm background transitions from
$T_{21}=0$~mK at very high redshift to $T_{21}\approx-100$~mK before
heating of the IGM by the first luminous sources, and then climbs to
$T_{21}\approx25$~mK at the beginning of the reionization epoch
before falling back to $T_{21}\approx0$~mK at the end of
reionization, the global mean redshifted 21~cm spectrum may contain
up to three relatively sharp features between
$50\lesssim\nu\lesssim200$~MHz that would not be represented well by
a power-law profile. For the large solid angles of a single antenna
beam, the mean redshifted 21~cm signal should vary little from one
location to another on the sky. \citet{2006MNRAS.371..867F} and
\cite{2004ApJ...608..611G} have calculated example global mean
redshifted 21~cm spectra for various assumptions of stellar formation
histories.

The specific approach employed with EDGES to exploit these expected
differences in spectral characteristics in order to overcome the
difficulty in separating the foreground and signal contributions in
the measured spectrum is to limit the scope of the experiment to test
for discontinuous features in the spectrum, since these would
necessarily be due to the rapid transitions in the redshifted 21~cm
brightness temperature and not the spectrally smooth foregrounds. In
particular, the frequency response of the system is designed to test
for fast reionization only (and not the transitions that might arise
at higher redshifts from cooling and heating of the IGM). In the
extreme case that the transition from a fully neutral to a fully
ionized IGM was virtually instantaneous, such that
$\dot{\bar{x}}_{HI}(z_r)\rightarrow\infty$, where $z_r$ is the
redshift of reionization, the contribution to the global spectrum at
the frequencies corresponding to the reionization epoch would
approach a step function.  A sharp feature resembling a step function
that is superimposed on the smooth power-law-like foreground spectrum
should be relatively easy to identify. And if reionization were to
progress more slowly, producing a smooth transition that spanned a
large range of redshifts and many tens of MHz, a simple model
could be used to set limits on the maximum rate of the transition.

\begin{figure}
\includegraphics[width=20pc]{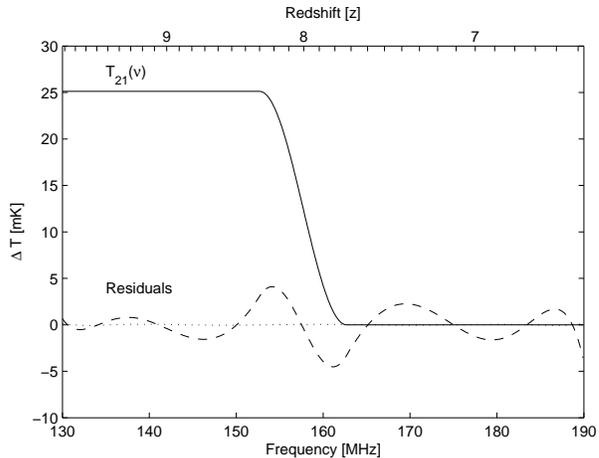}
\caption[Model sky spectrum used for EOR constraints]{
\label{f_edges_model} Example of redshifted 21~cm contribution
(solid) to $T_{sky}$ based on the model described in
\S~\ref{s_edges_limits} with $\Delta T_{21}=25$~mK, $z_r = 8$, and
$\Delta z = 0.6$. The residuals are also shown for a
seventh-order polynomial fit to a simulated spectrum
between 130 and 190~MHz with (dash) and without (dot) the redshifted 21 cm contribution.  The foreground contribution was modeled for the plot using $\beta=2.5$, and $T_{gal}(150$ MHz$) = 250$~K.}
\end{figure}

In principle, a variety of such models could be devised to use in
tests for the presence of a step feature in the radio spectrum due to
a rapid reionization. A simple low-order polynomial fit to the
measured spectrum would reveal such a discontinuous feature in the
residual spectrum after subtracting the fit and, thus, would be able
to determine the redshift range of a rapid reionization. Figure
\ref{f_edges_model} illustrates this approach by plotting a model
(described in Section \ref{s_edges_limits}) of the redshifted 21 cm
contribution to the measured spectrum along with the residuals after
a seventh-order polynomial fit is removed from a simulated sky
spectrum. This is the method used for the preliminary EDGES
measurements. An advantage of this approach for global reionization
experiments is that, given sufficient sensitivity, even a null result
could still constrain $\dot{\bar{x}}_{HI}(z)$ and, thereby,
distinguish between slow and fast reionization scenarios.

\section{Experiment Design}
\label{s_edges_system}

By focusing (at least initially) on confirming or ruling out a fast
reionization scenario, the design of the EDGES system is able to be
relatively simple.  The primary need is to reduce any instrumental or
systematic contributions to the measured power spectrum that vary
rapidly with frequency, since these could be confused with a sharp
feature in the spectrum due to a fast reionization of the IGM. Such
contributions could be due to terrestrial transmitters, reflections
of receiver noise or sky noise from nearby objects, undesirable
resonances within the electronics or the radio-frequency interference
(RFI) enclosures, or spurious signals introduced by the digital
sampling system. In this section, we provide an overview of the
experimental design and setup, highlighting aspects that are relevant
to reducing the effects of both the external and internal sources of
systematic errors. Additional details on the analysis of systematic
contributions and the hardware design can be found in the EDGES
memorandum
series\footnote{http://www.haystack.mit.edu/ast/arrays/Edges/}.

\subsection{Site Selection}

Some of the contributions to the systematic uncertainty listed above
can be addressed by careful selection of the observing site. Avoiding
terrestrial transmitters (primarily from FM radio and television
stations) is the most serious problem.  Even at
distances of hundreds or thousands of kilometers, tropospheric
ducting and scattering (troposcatter), sporadic E propagation in the
ionosphere, and reflections from meteors are all capable of
transferring a significant amount of power from Earth-based
transmitters. The background produced by the integrated effect of
many distant transmitters may have significant spectral structure
above the expected redshifted 21 cm level.  For example, a single,
100 kW FM radio station at 300 km from the observing site could
produce up to a 100 K effective temperature in a 1 MHz channel due to
troposcatter, or 100~mK due to meteor reflections. Fortunately, these
mechanisms of atmospheric propagation exhibit diurnal or transient
behavior (as is the case for sporadic E propagation, tropospheric
ducting, and meteor reflections) or require specific geometries for
peak efficiency (as is the case for troposcatter), making sensitive
measurements possible at remote sites at least some of the time.

Another concern is that local objects in the environment will scatter both external noise and receiver noise, which will then be picked up by the system and correlate with the original noise, causing sinusoidal ripples in the measured spectrum. We have estimated the magnitude of the reflections of the Galactic foreground from objects like trees and mountains on the horizon where the antenna gain is reduced by a factor of 20 dB or more. As long as objects subtend solid angles under about 100~deg$^2$, the spectrum will only be affected by a few parts per million (\textit{ppm}). We have also considered the magnitude of noise originating from the receiver that will be returned by a nearby scatterer. Even if we assume that this noise is perfectly correlated with the internal receiver noise, it will only produce ripples in the spectrum at the level of a few \textit{ppm} provided that the object, like a tree subtending a few deg$^2$, is more than $\sim100$~m away or a larger object, subtending $\sim100$~deg$^2$ is more than $\sim1$~km away.

Reflections of signals from compact radio sources may also be correlated. In this case the scatterer and the receiving antenna act like an adding interferometer to produce ripples in the spectrum. However, these effects are extremely small since a 1~Jy source results in under 1~mK of antenna temperature for the dipole-based EDGES system and the reflected signal is much smaller still. The ground reflection has been eliminated by placing the antenna on the ground. A brief discussion on the impact of these effects in radio astronomy measurements can be found also in \citet{rohlfs_wilson}.

\subsection{Hardware Configuration}
\label{s_edges_config}

Following a careful choice of the deployment site for the experiment, the remaining sources of systematic uncertainty result from the hardware design of the system.  The EDGES system consists of three primary modules: 1) an antenna, 2)
an amplifier and comparison switching module, and 3) an
analog-to-digital conversion and storage unit.  The antenna, shown in
Figure~\ref{f_edges_photos}, is a ``fat'' dipole-based design derived
from the four-point antenna of \cite{fourpoint1, fourpoint2}.  The
design was chosen for its simplicity and its relatively broad
frequency response that spans approximately an octave.  The response
of the antenna was tuned to 100 to 200~MHz by careful selection of
the dipole dimensions. In order to eliminate reflections from the
ground and to reduce gain toward the horizon, the antenna is placed
over a conducting mesh that rests directly on the ground. The mesh is
constructed from thin, perforated metal sheets to reduce weight and
is shaped to match an octagonal support structure below the ground
screen.  The diameter of this ground screen is approximately 2~m.

Although the antenna is constructed with perpendicular dipoles
capable of receiving dual linear polarizations, only one polarization
of the crossed-dipole is sampled by the receiver in order to reduce
the cost of the system.  This is acceptable since the spatially
averaged all-sky spectrum is expected to have essentially no
polarized component. The Galactic foreground does exhibit strong
polarization in certain regions, such as the ``fan region'' around
$\ell\approx140^\circ$, $b\approx8^\circ$, which has an extended
polarized component of about 3~K \citep{1973MNRAS.163..147W}. Such a
region could produce a ripple in the measured spectrum from a single
linear polarization as the polarization angle rotates with frequency.
Under the worst circumstances, if such a region were located at the
peak of the EDGES beam, the magnitude of the ripple could reach
$\sim50$~mK. Away from the Galactic plane, however, where EDGES
observations are generally targeted in order to reduce the system
temperature, it is more likely that the effects of polarization would
be at least an order of magnitude lower. Furthermore, if the rotation
measure (RM) is of order 10 rad m$^{-2}$, then the polarized
component could be averaged out over $\sim1$~MHz. Nevertheless, in
future versions of EDGES, both ports of the antenna will be sampled
in order to check for polarization effects and other systematic
effects that result from the non-uniformity of Galactic radiation.

A dipole antenna is naturally a balanced electrical system. To
convert from the balanced antenna leads to the unbalanced receiver
system (in which one lead is grounded), a short coaxial cable
enclosed in a clamp-on split ferrite core with a high impedance is
used as a common-mode choke balun\footnote{This balun provides a 1:1
impedance transition and operates on the same principle as the
quarter wavelength sleeve balun described by \citet[page
742]{KrausAntennas}. The ferrite provides a high impedance over a
wide frequency range to reduce the common-mode currents, whereas the
sleeve balun provides a high impedance over only a limited frequency
range close to the quarter wavelength resonance.} and is connected
directly to the terminals of the antenna with the central conductor
fastened to one element and the braided shielding to the other.

The amplifier module consists of two stages that are contained in
separate aluminum enclosures to reduce coupling between the low-noise
amplifiers. Each stage provides 33~dB of gain for a total of 66~dB.
Bandpass filtering of the signal is also performed in the second
stage, and the resulting half-power bandwidth spans approximately 50
to 330~MHz. The amplifier chain can be connected through a voltage
controlled three-position switch to one of three inputs: the antenna,
an ambient load, or an ambient load plus a calibration noise source.
Switching between the ambient load and the antenna provides a
comparison to subtract spurious instrumental signals in the measured
sky spectrum.

Impedance mismatch between the antenna and the amplifiers causes
reflections of the sky noise within the electrical path of the
instrument that produce an undesirable sinusoidal ripple in the
measured spectrum due to the frequency-dependence of the phase of the
reflections at the input to the amplifier. To reduce the effects of
these reflections in EOR measurements, the input to the amplifier
chain is connected directly to the balun on the antenna (with no
intermediate transmission cable), as shown in
Figure~\ref{f_edges_photos}. While absolute calibration is limited in
this configuration by the effect of the unknown phases of the
reflections on the measured spectrum, the compact size of the antenna
and the small signal path delays result in a smooth spectral
response.

The amplifier module is connected to the analog-to-digital conversion
module by three low-loss coaxial transmission cables.  The cables
provide power, switching control, and signal transmission,
respectively. Common-mode current on these cables (i.e. current that
is on the outer surface of the shielding in the coaxial cable, or
current that is unidirectional on both the central conductor and
inner surface of the shielding) is also capable of producing
reflections and additional sinusoidal ripples in the measured
spectrum. The ferrite core balun used between the antenna and
amplifiers allows common-mode current of approximately 10\% of the
differential mode. Although most of this current is transferred to
the ground screen by direct contact between the amplifier module
casing and the ground screen, some current persists and leaks through
the casing of the amplifier module and onto the shielding of the
three cables connecting the amplification module to the
analog-to-digital conversion module. To reduce this current to less
than 0.005\% of the differential mode current, additional clamp-on
ferrite cores are placed every meter on the transmission cables.

Finally, the analog-to-digital conversion is accomplished with an
Acqiris
AC240\footnote{http://www.acqiris.com/products/analyzers/cpci-signal-analyzers/ac240-platform.html}
8-bit digitizer with maximum dynamic range of 48~dB (although, in
practice, the effective dynamic range was substantially lower due to
coupling between the digital output of the converter and its input).
The AC240 uses an embedded field programmable gate array (FPGA) to
perform onboard Fast Fourier Transform (FFT) and integration of the
power spectrum in realtime. The spectrometer is clocked at 1~GS/s and
the Fourier transform processes 16,384 channels, giving a bandwidth
of 500~MHz and a raw spectral resolution of about 30~kHz. The
broadband spectrometer employs the FPGA code of
\citet{2005A&A...442..767B} and a Blackman-Harris window function is
used to improve the isolation between neighboring frequency channels
at the expense of reducing the effective spectral resolution to
122~kHz. The unit is contained on a CompactPCI card connected to a
host computer. The digitizer and host computer, along with a power
transformer and interface circuitry for controlling the amplifier
module with the serial port of the computer, are enclosed in an
aluminum box to prevent self interference.

\subsection{The Measured Spectrum}
\label{s_data_acquisition}

\begin{figure}
\includegraphics[width=22pc]{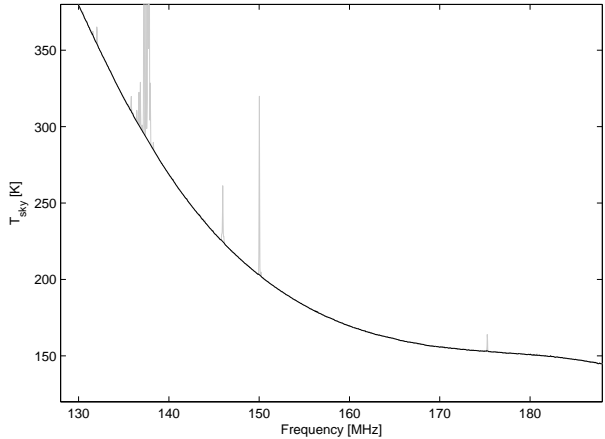}
\caption[Measured sky spectrum used for EOR constraints]{
\label{f_edges_spectrum} Integrated spectrum used for upper limit
analysis of reionization signal.  The sky temperature, $T_{sky}$, is
an estimate based on modeled values for cable losses and no
correction for antenna reflections. The spectrum represents the best
10\% of the data from observations over two nights.  It is selected
by discarding individual observation cycles (see
\S~\ref{s_data_acquisition}) containing periods of particularly
intense radio frequency interference.  A total of approximately 1.5 h
of integration is included (3.75 h including the ambient load and
calibrator noise source measurements in each cycle). The black curve
shows the spectrum after de-weighting the interferers (shown in gray)
present in the retained observations.}
\end{figure}

The measured spectra from each of the three switch
positions can be combined to produce a calibrated estimate of the
true sky spectrum. The three spectra are given by
\begin{equation}
\begin{array}{ccl}
p_0 & = & g ~ (T_L + T_R) ~ (1 + n_0) \\
p_1 & = & g ~ (T_L + T_R + T_{cal}) ~ (1 + n_1) \\
p_2 & = & g ~ (T_A + T_R) ~ (1 + n_2)
\end{array}
\end{equation}
where the explicit frequency dependence of each term has been dropped
and $p_0$ is the spectrum for the ambient load, $p_1$ is the spectrum
for the ambient load plus calibration noise, and $p_2$ is the
spectrum for the antenna.  In this terminology, $g$ is the gain,
$T_L$ is the ambient load temperature, $T_R$ is the receiver noise
temperature, $T_{cal}$ is the calibration noise temperature, and
$T_A$ is the antenna temperature. Thermal uncertainty in the
measurements is explicitly included in the Gaussian random variables
$n_0$, $n_1$, and $n_2$, the magnitudes of which are given by $n_i =
(\epsilon~b~\tau_i)^{-1/2}$, where $\epsilon=0.5$ is the efficiency
for the Blackman-Harris window function (which could be improved to
0.93 by processing two sets of overlapping windows), $b =
122\times10^3$~Hz is the resolution bandwidth, and $\tau_i$ is the
integration time in seconds (for each switch position, $i$).
Temporarily setting the noise terms to zero, $n_i\rightarrow0$, and
solving for the antenna temperature yields
\begin{equation}
T_A = T_{cal}\frac{p_2 - p_0}{p_1 - p_0} + T_L.\label{eqn_ta}
\end{equation}

In practice, the impedance match between the antenna and receiver is
not perfect and some of the incident sky noise may be reflected back
out of the system.  This produces deviations between the derived sky
temperature found using Equation~\ref{eqn_ta} and the true sky
spectrum. Independent measurements of the impedance mismatch can be
used to correct these deviations by applying a frequency-dependent
multiplicative factor to $T_A$ that is proportional to the inverse of
the reflection coefficient. For the EDGES system, this correction was
measured by two methods: we used a network analyzer in the laboratory
to determine the impedance of the antenna, and we reconfigured the
system in the field with a long cable inserted between the antenna
and amplifier module so that reflections between the two elements
were visible in the measured spectrum and could be used to calibrate
the reflection coefficient. In both sets of measurements, the
corrections were found to be small (of order 1\%) and smooth (able to
be fit by a low-order polynomial in frequency) over the band of
interest. For the remainder of this paper, we will ignore this correction since its effects are easily absorbed by the polynomial fit technique used to constrain the redshifted 21 cm contribution to the spectrum.

Adding the noise terms back in and solving in the limit that $T_{cal}
\gg (T_L \approx T_A) > T_R$ results in an estimate of the thermal
uncertainty per frequency channel of approximately
\begin{equation}
\Delta T_{A,rms} \approx \sqrt{ n^2_0 (T_L + T_R)^2 + n^2_1 (T_L)^2 +
n_2^2(T_A+T_R)^2}.
\end{equation}
For optimal efficiency, the three terms contributing to $\Delta
T_{A,rms}$ should be comparable in magnitude.  Substituting
$T_L=300$~K, $T_A=250$~K and $T_R=20$~K, we find that the terms are
comparable as long as approximately equal time is spent in each
switch position.  In addition, a 1~hour integration in each switch
position (3~hours total) will result in a thermal uncertainty in the
estimate of the antenna temperature of $\Delta T_{A,rms} \approx
35$~mK within each 122~kHz frequency channel.

To acquire a series of estimates of the true sky spectrum using this
technique, software on the host computer cycles the amplifier module
between the three switch positions and triggers the digitizer to
acquire, Fourier transform, and accumulate data for a predefined
duration at each of the switch positions. The integration durations
per switch position are $\tau_{\{0,1,2\}}=\{10, 5, 10\}$ seconds for
the ambient load, ambient load plus calibration noise source, and
antenna, respectively, giving a duty cycle of about 40\% on the
antenna.  This loop is repeated approximately every 25 seconds for
the duration of the observations and the resulting measurements are
recorded to disk.

\section{Initial Results}
\label{s_edges_results}

\begin{figure}
\includegraphics[width=22pc]{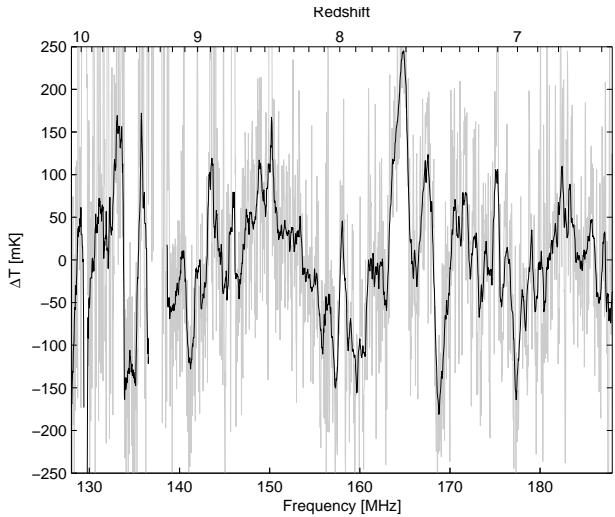}
\caption[Instrumental residuals in measured spectrum]{
\label{f_edges_residuals} Residuals after subtraction of
seventh-order polynomial fit to measured spectrum shown in Figure
\ref{f_edges_spectrum}. The gray line is the raw spectrum with 122
kHz resolution. The black line is after smoothing to 2.5 MHz
resolution to reduce the thermal noise to below the systematic noise.
The \textit{rms} of the smoothed fluctuations is approximately 75~mK
(see Figure \ref{f_edges_rms_vs_time}).}
\end{figure}

The EDGES system was deployed at the radio-quiet Mileura Station in
Western Australia from 29~November through 8~December~2006.  These
dates were chosen such that the Galactic center would be below the
horizon during most of the night, keeping the system temperature as
low as possible for the measurements.  The system was located
approximately 100~m from the nearest buildings in a clearing with no
nearby objects and no obstructions above $\sim5^{\circ}$ on the
horizon, and the antenna was aligned in an approximately
north-south/east-west configuration. The system was operated on 8
consecutive nights during the deployment, with 5 of the nights
dedicated to EOR observing. In total, over 30~h of relevant drift
scans were obtained, but strong, intermittent interference from
satellites complicated the measurements and only approximately 8~h of
high-quality observations were retained as the primary data set.
Although the satellite interferers that complicated the measurements
were narrow-band and, in many cases, were easily removed through
excision of the effected spectral channels, the limited dynamic range
of the EDGES system resulted in clipping of the analog-to-digital
converter and corruption of the full band during especially strong
transmissions.  This required all channels to be omitted from the
data set in those instances. In particular, it was found that the low
Earth orbiting satellites of Orbcomm (transmitting between
approximately 136 and 138 MHz), as well as satellite beacons (at 150
MHz) from discarded spacecraft were particularly troublesome. The
Orbcomm activity was somewhat variable and usually decreased during
the night.  The typical duration of a pass was approximately 15
minutes, during which time the power in the satellite signal could
easily reach an order of magnitude greater than the integrated sky
noise over the band. While previous observations at the site with
prototype MWA equipment \citep{2007AJ....133.1505B} have demonstrated
(in a subset of the full target band) that it is possible to reach
the sensitivities required for EDGES despite the satellites and other
sources of interference, improvements to the EDGES digitizing system,
such as an upgrade of the analog-to-digital converter to 10 or 12
bits, would certainly help to alleviate the difficulties encountered
during this observing campaign and increase the usable fraction of
measurements.

From the primary data set remaining after transient RFI exclusion, a
stringent filter was applied to select the best 1.5~h of sky-time
when transient interference signals were weakest. The final cut of
data included measurements from multiple nights and spanned a range
of local apparent sidereal time (LST) between 0 and 5~h. The sky
temperature at 150~MHz derived from the system during this period was
found to have a minimum of $\sim240$~K at about 3~h LST and a maximum
of $\sim280$~K at 5~h LST. The integrated spectrum generated from
these measurements is shown in Figure~\ref{f_edges_spectrum}.
Frequency channels containing RFI were identified in the integrated
spectrum by an algorithm that employs a sliding local second-order
polynomial fit and iteratively removes channels with large errors
until the fit converges.  The affected channels were then weighted to
zero in subsequent analysis steps. To look for small deviations from
the smooth foreground spectrum, a seventh-order polynomial was fit to
the measured spectrum between 130 and 190~MHz (where the impedance
match between the antenna and receiver was nearly ideal) and
subtracted.

\begin{figure}
\includegraphics[width=22pc]{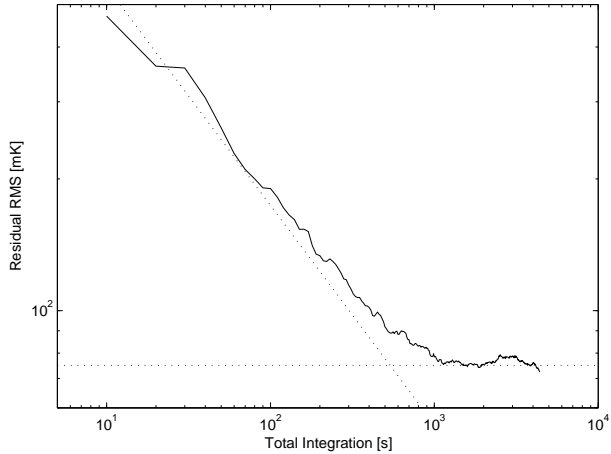}
\caption[Amplitude of residuals as a function of integration time]{
\label{f_edges_rms_vs_time} Characteristic amplitude of the residuals
to the polynomial fit as a function of integration time on the sky.
The \textit{rms} follows a thermal $(b \tau)^{-1/2}$ dependency until
saturating at a constant 75~mK noise level due to the instrumental
errors introduced into the measured spectrum.  The dotted lines are
guides for the eye showing a $(b \tau)^{-1/2}$ profile and a constant
75~mK contribution.}
\end{figure}

The residual deviations in the measured sky spectrum after the
polynomial fit and subtraction are shown in
Figure~\ref{f_edges_residuals}. The \textit{rms} level of systematic
contributions to the measured spectrum was found to be $\Delta
T_{rms} \approx 75$~mK, a factor of $\sim3$ larger than the maximum
expected redshifted 21~cm feature that would result from a rapid
reionization. Although it is not obvious in
Figure~\ref{f_edges_residuals}, the variations in the residuals are
due to instrumental contributions and not thermal noise. The large
variations between 163 and 170~MHz are due to the 166 MHz PCI-bus
clock of the AC240 and computer, while the gap centered at
approximately 137~MHz is due to RFI excision of the Orbcomm satellite
transmissions over a region spanning more than 2.5~MHz.  Analysis of
the dependence of $\Delta T_{rms}$ on integration duration is shown
in Figure~\ref{f_edges_rms_vs_time} and illustrates that the
\textit{rms} of the residuals follows a thermal profile $\sim(b
\tau)^{-1/2}$ initially and then saturates to a constant value. After
smoothing to 2.5~MHz resolution ($\Delta z\approx0.2$), the
instrumentally dominated 75~mK threshold is reached in approximately
20~minutes (1200~s) of integration on the sky (50~minutes of total
integration in all three switch positions). Reordering the individual
25-second observation cycles used in the full integration does not
change the behavior in Figure~\ref{f_edges_rms_vs_time}, and longer
integrations (up to approximately 3 h of sky time), using observation
cycles with more intense interference, continued to decrease the
thermal noise, but leave the spurious signals and systematic effects
unchanged.

\subsection{Limits on Reionization History}
\label{s_edges_limits}

Although the sensitivity level of the initial observations with the
EDGES system was limited by instrumental effects in the measured
spectrum at a level greater than the expected maximum contribution
due to redshifted 21~cm emission, weak constraints can still be
placed on the reionization history.  In addition, it is possible to
make a quantitative assessment of how much improvement must be made
before significant constraints are possible, as well as to
characterize the best-case outcome of future efforts using similar
approaches. To begin, we introduce a model for the sky spectrum such
that
\begin{equation}
\label{eqn_sky_temp} T_{sky}(\nu) = T_{gal}(\nu) + T_{cmb} +
T_{21}(\nu)
\end{equation}
where $T_{gal}$ represents the contribution of all the foregrounds
(and is dominated by the Galactic synchrotron radiation),
$T_{cmb}=2.73$~K is the CMB contribution, and $T_{21}$ is the
specific form for the frequency-dependence of the redshifted 21~cm
emission during the transition from the fully neutral to fully
ionized IGM. This model neglects any directional or temporal
variation in $T_{sky}$ and, therefore, implicitly assumes an angular
average over the antenna beam and a time average over the drift scan
measurements performed for the experiment. Since $T_{cmb}$ and
$T_{21}$ are taken to be constant over the sky, only the $T_{gal}$
contribution is affected by this simplification.  This does not
impact the result, however, as long as the foreground emission varies
slowly on the sky and the antenna pattern changes slowly with
frequency---conditions that are presumed to be met in the high
Galactic latitude region sampled by the dipole-based EDGES system. As
a test of this assumption, we calculated the residuals after the
polynomial fit for a bright source with flux comparable to Cas A
(1400~Jy at 100~MHz) and spectral index $\beta=2.77$ at various
positions in the antenna beam using simulated beam patterns to
determine the frequency-dependence. We found, in all cases, less than
a $\sim50$~$\mu$K residual.

During the reionization epoch, we define $T_{21}$ to be given by
\begin{equation}
T_{21}(z) = \Delta T_{21} \frac{1}{2} \left \{ 1 + cos \left [ \frac{
\pi (z_r - z - \Delta z / 2)} { \Delta z} \right ] \right \},
\end{equation}
where $\Delta T_{21}$ is constant and is the maximum amplitude of the
redshifted 21~cm contribution, $z_r$ is the redshift when
$\bar{x}_{HI}(z_r)=0.5$, $\Delta z$ is the total duration of the
reionization epoch, and we use $\nu = 1420 / (1+z)$~MHz to convert
back to frequency units. Before the reionization epoch ($z>z_r+\Delta
z/2$), $T_{21} \equiv \Delta T_{21}$, while after reionization
($z<z_r-\Delta z/2$), $T_{21}\equiv0$. The exact form of the
transition used for $T_{21}$ has little influence on the outcome of
the constraints as long as it is reasonably smooth.
Figure~\ref{f_edges_model} illustrates the modelled redshifted 21~cm
spectrum.  The free parameters in the model are $z_r$, $\Delta z$,
and $\Delta T_{21}$.

\begin{figure}
\includegraphics[width=20pc]{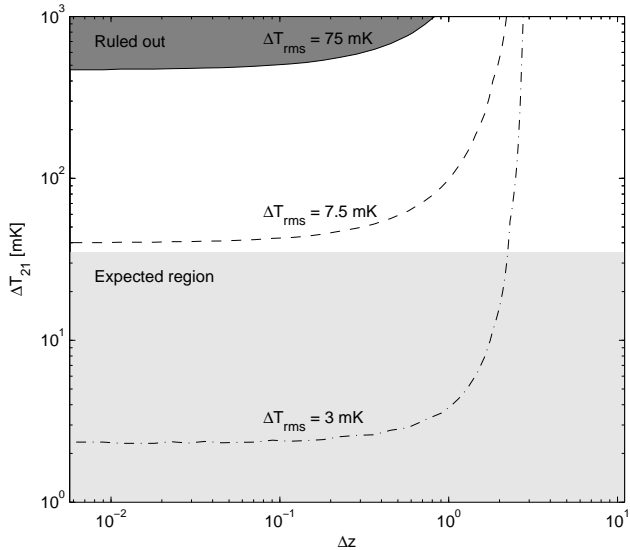}
\caption[Initial constraints by EDGES on reionization history]{
\label{f_edges_constraint} Constraints placed by EDGES on the
redshifted 21~cm contribution to the sky spectrum. The dark-gray
region at the top-left is the portion of the parameter space ruled
out by the initial EDGES results with $\Delta T_{rms}=75$~mK (solid
line). The dashed lined labelled $\Delta T_{rms}=7.5$~mK and the
dotted line labelled $\Delta T_{rms}=3$~mK indicated the constraints
that could be placed on reionization if the experimental systematics
were lowered to the respective values.  The light-gray region along
the bottom is the general range of parameters believed to be viable.
The redshifted 21~cm contribution to the spectrum is modelled
according to the description in \S~\ref{s_edges_limits} with $z_r =
8$.}
\end{figure}

For the EDGES best-response frequency range, a center redshift around
$z_r=8$ allows the largest range of $\Delta z$ to be explored.  By
simulating the combined sky spectrum, $T_{sky}$, for a range of the
two remaining free parameters, we can determine the \textit{rms} of
the residuals that would remain following the polynomial fit used in
the EDGES data analysis.  Comparing the \textit{rms} of the residuals
in the models to the 75~mK \textit{rms} of the initial measurements
gives a good estimate of the region of parameter space ruled out so
far. Figure~\ref{f_edges_constraint} illustrates the results of this
process. The line defining the ruled-out region is computed by
finding the locus of parameters that make the \textit{rms} residuals
in the model equal to 75 mK. While a more statistically robust
analysis is clearly possible, little benefit would be gained for the
initial measurements presented here due to the severe systematic
effects present in the spectrum.

From Figure~\ref{f_edges_constraint}, it is clear that the initial
results constrain only a small portion of parameter space that is
well outside the expected region for both the intensity of the
redshifted 21~cm signal and the duration of reionization. The best
constraint, in the case of a nearly instantaneous reionization, is
that the redshifted 21~cm contribution to the spectrum is not greater
than about $\Delta T_{21} \lesssim 450$~mK before the transition.
Reducing the systematic contributions in the measured spectrum by
more than an order of magnitude to $\Delta T_{rms}<7.5$~mK would
begin to allow meaningful constraints, while an improvement of a
factor of 25 to $\Delta T_{rms} \approx 3$~mK would be able to rule
out a significant portion of the viable parameter range and constrain
$\Delta z > 2$. In principle, such an improvement is possible with
minor modifications to the EDGES system. Reaching a systematic
uncertainty below $\sim3$~mK, however, is likely to be infeasible
without a redesign of the experimental approach because errors in the
polynomial fit to the overall power-law-like shape of the sky
spectrum, $T_{sky}(\nu)$, are the dominant source of uncertainty
below that level in the current approach.  The sharp cut-off at
$\Delta z\approx2$ in parameter space is the result of using a
seventh-order polynomial to fit a 60~MHz bandwidth, thus yielding a
maximum residual scale size of order 10~MHz, which corresponds to
$\Delta z\approx2$ at $z\approx8$.  If the same polynomial could be
reasonably fit to a larger bandwidth (or a lower-order polynomial fit
to the existing bandwidth), then $\Delta z$ could be probed to larger
values.

\section{Conclusion}

In principle, useful measurements of the redshifted 21~cm background
can be carried out with a small radio telescope. These measurements
would be fundamental to understanding the evolution of the IGM and
the EOR. In particular, the global evolution of the mean spin
temperature and mean ionization fraction of neutral hydrogen in the
high redshift IGM could be constrained by very compact instruments
employing individual radio antennas.  We have reported preliminary
results to probe the reionization epoch based on this approach from
the first observing campaign with the EDGES system. These
observations were limited by systematic effects that were an order of
magnitude larger than the anticipated signal and, thus, ruled out
only an already unlikely range of parameter space for the
differential amplitude of the redshifted 21~cm brightness temperature
and for the duration of reionization. Nevertheless, the results of
this experiment indicate the viability of the simple global spectrum
approach.

Building on the experiences of these initial efforts, modifications
to the EDGES system are underway to reduce the residual systematic
contribution in the measured spectrum and to expand the frequency
coverage of the system down to 50~MHz or lower in order to place
constraints on the anticipated transition of the hyperfine line from
absorption to emission as the IGM warms before the EOR. Constraining
the redshift and intensity of this feature would be very valuable for
understanding the heating history of the IGM and, since the
transition has the potential to produce a step-like feature in the
redshifted 21~cm spectrum with a magnitude over 100~mK (up to a
factor of 4 larger than the amplitude of the step during the
reionization epoch), it may be easier to identify than the transition
from reionization---although the sky noise temperature due to the
Galactic synchrotron foreground increases significantly at the lower
frequencies, as well. Through these and other global spectrum
efforts, the first contribution to cosmic reionization science from
measurements of the redshifted 21~cm background will hopefully be
achieved in the near future.

\acknowledgements

This work was supported by the Massachusetts Institute of Technology,
School of Science, and by the NSF through grant AST-0457585.


\end{document}